\documentclass[useAMS,usenatbib,usegraphicx]{mn2e}
\usepackage[usenames,dvipsnames]{xcolor}
\usepackage{times}
\newcommand{\lae}{\lower 2pt \hbox{$\, \buildrel {\scriptstyle <}\over {\scriptstyle
\sim}\,$}}

\begin{document}
\title{Radio Emission from the Bow Shock of G2}
\author[P. Crumley and P. Kumar]{P.~Crumley,$^1$\thanks{E-mail:
    crumleyp@physics.utexas.edu, pk@surya.as.utexas.edu}
  P.~Kumar,$^2$\footnotemark[1]\\ $^1$Physics Department, University
  of Texas at Austin, Austin, TX 78712\\ $^2$Astronomy Department,
  University of Texas at Austin, Austin, TX 78712 }

\maketitle
\begin{abstract}
The radio flux from the synchrotron emission of electrons accelerated
in the forward bow shock of G2 is expected to peak when the forward
shock passes pericenter, possibly 7 to 9 months before the center of
mass of G2 reaches pericenter $\sim 3\times10^{15}\ $ cm from the
Galactic Center \citep{Narayan,Sadowski2,Sadowski}.  In this letter,
we calculate the radio emission from the forward and reverse shock if
G2 is a momentum-supported bow shock of a faint star with a high
mass-loss rate as suggested by \citet{Scoville, Ballone}. We show that
the radio flux lies well below the quiescent radio flux of Sgr A* and
will be difficult to detect. By contrast, in the cloud model of G2,
the radio flux of the forward shock is predicted to be much larger
than the quiescent radio flux and therefore should have already been
detected \citep{Narayan, Sadowski}. Therefore, radio measurements can
reveal the nature of G2 well before G2 completes its periapsis
passage.
\end{abstract}
\begin{keywords}
galactic center, black hole physics, radiation mechanism, non-thermal
\end{keywords}

\section{Introduction}
G2, a recently discovered spatially-extended red source, is on a
nearly radial orbit headed towards the $M\sim 4.31\times
10^6\ M_\odot$
($R_s=2GM/c^2\sim 1.27\times10^{12}\ {\rm cm}$), 
supermassive black hole at the Galactic center, Sgr A*
\citep{Gillessen}.  As G2 plunges towards Sgr A* at a supersonic
speed, it will drive a bow shock into the hot interstellar medium
(ISM). As electrons cross the forward shock, they will be accelerated
and emit synchrotron radiation. The radio synchrotron emission from
the forward shock has been predicted to be observable by
\citet{Narayan, Sadowski}. \citet{Sadowski2} predicted that the radio
emission should peak 7 to 9 months before the center of mass crosses
periapsis in spring of 2014, reaching a distance of a mere $\sim
3\times10^{15}\ {\rm cm}\ (\sim 2400R_s)$ away from Sgr A*
\citep{Phifer,GillessenNewest}. This close encounter may give a
unique opportunity to probe the accretion disk of Sgr A* and the ISM
near periapsis by measuring the radio flux of the forward bow shock.
The radio flux from the forward shock has yet to be observed;
depending on the orbital parameters of G2, it may not be expected to
peak until late summer or early autumn of 2013, although perhaps $\sim
10\%$ of G2 has already passed periapsis \citep{GillessenNewest}. In
this letter, we show that the expected radio flux from the forward
shock is model dependent, suggesting an intriguing possibility: the
radio flux from the forward shock may reveal the nature of G2.

The nature of G2 is undetermined. When first discovered,
\citet{Gillessen, GillessenNew} hypothesized that G2 was a
pressure-confined, non-self-gravitating gas cloud, due to the fact
that the Brackett-gamma (Br-$\gamma$) luminosity of G2 is not changing
with time, $L_{\rm Br\gamma}\sim 2\times 10^{-3} L_\odot$, and the
Br-$\gamma$ velocity dispersion is increasing in a manner that is well
fit by a gas cloud with a radius of $\sim 2\times10^{15}\ {\rm cm}$
being tidally sheared by Sgr A*. The compactness of G2 necessitates
that this gas cloud must have formed shortly before G2 was first
discovered in 2002. However, there is not a clear source of a gas
cloud in the region as the ISM at a distance $\sim
5\times10^{16}\ {\rm cm}$ from Sgr A* is not susceptible to thermal
instabilities. One possible source of a gas cloud is colliding stellar
winds \citep{Burkert,Schartmann}.  Alternatively, there
  is another class of models where G2 contains a very faint stellar
  core that emits gas as it falls towards Sgr A* \citep{clay,
    Scoville, Ballone}. The ionized gas is then tidally sheared and is
  the source of the Br-$\gamma$ radiation seen as G2. In
  \citet{Scoville}, the ionized gas that is the source of the
  Br-$\gamma$ radiation is located in the cold dense inner shock of a
  momentum-supported bow shock between a stellar wind from a hidden,
  young star and the hot ISM. \citet{Ballone} performed
  hydrodynamical simulations in which they evolved the stellar wind
  shock in the gravitational potential of Sgr A*, and they found that
  the $v_{lsr}$ dispersion observations are well-matched by a star
  with a mass-loss rate of $\dot{M}=8.8\times10^{-8}M_{\odot}/{\rm
    yr}$ and with a wind speed of $50\ {\rm km/s}$.  The ionization
  source of the inner shock in this model is not entirely clear, as
  the predicted Br-$\gamma$ radiation  does not perfectly match the
  observations. \citet{Scoville} proposed that hydrogen is
  collisionally ionized in the inner shock of the stellar wind,
  although they overestimated the ionization ability of the wind and
  underestimated the ionizing background from the O stars in the
  galactic center. \citet{Ballone} just assumed the entire inner shock
  was ionized, which leads to the Br-$\gamma$ flux increasing as G2
  approaches pericenter, which is not what has been observed. Future
  work will need to be done to understand the radiation mechanism in
  the stellar wind model of G2.

If G2 is a pressure-confined diffuse cloud, then it will likely be
destroyed during periapsis passage. If G2 is the inner bow shock of a
star it will likely survive \citep{Gillessen, Anninos, Schartmann,
  Scoville}. Additionally, if G2 is a cloud, how its size changes as it approaches
Sgr A* is not clear while the stellar wind model of G2 makes a strong
prediction about how the size of the forward bow shock evolves as G2
passes through periapsis. 

In the cloud model, as G2
  heads closer to Sgr A* it will be stretched in the longitudinal
  direction, compressed in the transverse direction by tidal forces,
  and compressed further by the increasing pressure of the ISM.  In
the 3-D simulation by \citet{Anninos}, for an isothermal cloud that
formed in 1995.5 with a radius of $1.875\times10^{15}\ {\rm cm}$, they
find that the cross sectional area of G2 one year before periapsis
shrunk by a factor of 4 to $\sim \pi\times 10^{30} {\rm cm^2}$. It
appears to be the same size at periapsis. The cloud is in the process
of being completely disrupted, so it is difficult to tell.

In this letter, we use the term ``cloud model'' to mean a model where
the cross section of G2 is approximately constant as it passes through
periapsis as in \citet{Narayan,Sadowski}.  In the stellar wind model,
far away from Sgr A* the size of the shock is approximately the same
as in the cloud model, but when G2 gets close to periapsis, its size
shrinks significantly, inversely proportional to the increase of the
pressure of the ISM. In other words, the area is proportional to the
distance from Sgr A* squared (see eq \ref{eq:areaEst}). The hydrodynamic
simulations in \citet{Ballone} show a similar decrease to the shocked
wind area. The decrease in area at periapsis drastically reduces the
expected radio flux.

G2 appears to be an extended object with the compact head structure
called G2, and a less dense and larger tail-like structure, G2t
\citep{GillessenNewest}. This paper focuses solely on the over-dense
head, G2, as the origin of G2t is not well known. If G2 is a diffuse
cloud then G2t is either material that was stripped from G2 as it fell
towards Sgr A*, or it was somehow formed through a similar process as
G2. If G2 has a stellar core emitting a wind, the tail may have been
from the outflow of the star when it was closer to apocenter.

In this letter, we extend the work of \citet{Narayan, Sadowski} to the
stellar wind model of G2. We apply \citet{Wilkin} analytic solution
of the momentum-supported bow shock model proposed by
\citet{Baranov} to estimate the size of the bow shock of
G2. Analytical calculations are better suited to estimate the forward
shock size than a hydrodynamic simulation such as \citet{Ballone}
because of stability issues in the simulation. The ISM surrounding Sgr
A* is convectively unstable. Previous simulations \citep[with the
  exception of][]{Sadowski2} dealt with this issue by evolving a
passive tracer field along with G2 and resetting the ISM to its
equilibrium value whenever a cell did not have a sufficient ratio of
tracer particles to ISM particles \citep{Anninos, Schartmann}.
As noted in \citet{Ballone}, this stabilization technique 
suppresses the forward shock in the ISM, making these simulations
incapable of properly resolving the forward shock.
Analytical calculations also allow us to predict how the radio flux
depends on the undetermined parameters of the stellar wind.

We use the geometry of the bow shock to calculate the expected
synchrotron flux in the radio band of both the outer forward shock and
the inner termination shock for the stellar wind model of G2. The
predicted flux of the forward shock is roughly two orders of magnitude
less than the flux predicted by \citet{Sadowski}, and lies an order of
magnitude below the quiescent radio emission of Sgr A* at 2 GHz. If
the stellar wind model is correct, G2 will likely not be observable in
radio frequencies. Therefore, a radio detection of G2 will shed light
on the nature of G2 well before periapsis passage.

In section \ref{sec:METHODS} of this letter, we briefly describe the
model we used for the ISM at the Galactic Center as well as the
geometry of the bow shock. In section \ref{sec:RESULTS} we show how
the expected peak synchrotron flux at 1.4 GHz and spectrum depend on
the nature of G2. In section \ref{sec:S2} we extend our
  results to the star S2. Finally, in section \ref{sec:CONCLUSIONS}
we briefly discuss our findings.

\section{Methodology}\label{sec:METHODS}
\subsection{Environment at the Galactic Center}
For the hot ISM environment surrounding Sgr A*, we adopt the same
dependence of density and temperature on distance $d$ from Sgr A* as
those used by \citet{Schartmann}, who used the model of
\citet{Yuan03}. However, the most recent \textit{Chandra} X-ray
observations of the Galactic Centre suggest that the radial density
profile may be flatter than the one used by Schartmann et al. \citep{Wang13}
\begin{equation}\label{eq:n_ISM}
n_{\rm ISM}=930\ {\rm cm^{-3}} 
\left(\frac{1.4\times10^4 R_s}{d}\right)
=1660\ d_{16}^{-1}\ {\rm cm^{-3}}
\end{equation}
\begin{equation}\label{eq:T_ISM}
T_{\rm ISM}=1.2\times10^8\ {\rm K} 
\left(\frac{1.4\times10^4 R_s}{d}\right)
=2.1\times10^{8} d_{16}^{-1}\ {\rm K}
\end{equation}
Throughout this paper the convention \(Q_x\equiv \frac{Q}{10^x}\) is
used, and unless otherwise noted all units are cgs.  To calculate the
distance and velocity of G2, we use the orbital parameters derived
from the Br-$\gamma$ observations given in
\citet{GillessenNewest}. However, for the purposes of illustration
G2's velocity is well matched by a free fall approximation, {\it i.e.}
$v_*\approx \sqrt{2GM/d}=3.4\times10^{8} d_{16}^{-0.5}\ {\rm
  cm/s}$. The Mach number is approximately $\mathcal{M}_*\approx
2$. Because of the approximation we made to the velocity, the Mach
number has no time dependence, but it will change with time when using
the proper elliptical motion of the orbit.

\subsection{Geometry of the Bow Shock}

\begin{figure}
\vspace{-.4cm}
\includegraphics[width=.47\textwidth]{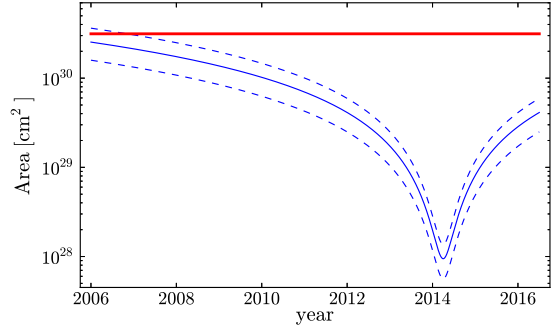}
  \caption{The cross section area of the forward shock as G2
    approaches periapsis. The solid blue line is the cross sectional
    area of the forward shock calculated in a bow shock correcting for
    the centrifugal force. The lower dashed blue line is the size of
    the forward shock if using the bow shock model of \citet{Dyson} as
    did \citet{Scoville}. The upper dashed blue line is
    $\pi\times\dot{M}v_w/(4\pi n k T)$, the cross-sectional area of
    the inner shock of G2, which agrees very well with the simulation
    by \citet{Ballone}. The thick red line is the area used in
    \citet{Sadowski}, $\pi\times10^{30}$ cm.}
\label{fig:area}
\end{figure}

To calculate the radio flux from the forward bow shock of G2, we
need to determine the area of its cylindrical cross section. We use
the same parameters for the isothermal stellar wind as in
\citet{Ballone}: a star with a mass-loss rate of $\dot{M}_* = 8.8
\times10^{-8} M_{\odot}/{\rm yr}$, a constant wind speed of
$v_w=50\ {\rm km/s}$, and a temperature of $T_w=10^4\ {\rm K}$. 
We include the dependence of our results on the stellar wind
parameters as they are not precisely constrained by observations.
The wind is supersonic, with a Mach number of $\mathcal{M}_w =
4.26\ T_4^{-0.5}(v_w/50\ {\rm km/s})$. To calculate the properties of
a stellar wind bow shock, we used the equations given in
\citet{Baranov} instead of \citet{Dyson} because \citet{Baranov}
correctly accounts for the centrifugal force due to the fluid moving
in a curved path inside of the shock. The effect of the centrifugal
force is to produce a larger bow shock, {\it e.g.}  \citet{Baranov},
than the solution given in \citet{Dyson} (see figure \ref{fig:area}
for a comparison). The stagnation radius, $r_0$, is the same in both
solutions and it occurs where the two ram pressures equal one another,
{\it i.e.} $\rho_w v_w^2=\rho_{\rm ISM} v_*^2$. For the adopted
parameters of $\dot{M}_*$ and $v_w$,
\begin{equation}\label{eq:rstag}
r_0  = \sqrt{\frac{\dot{M}_*v_w}{4\pi m_p n_{\rm ISM}v_*^2}}
\approx   8\times10^{13}\ d_{16}
{\textstyle 
\left(\frac{\dot{M}_*v_w}{2.8\times10^{25}\ {\rm dyn}}\right)^{0.5}
{\rm cm}}
\end{equation}
The shock is axisymmetric about the stellar velocity vector, and the
distance of the bow-shock surface from the star depends on the
azimuthal angle measured from the stagnation point, $\varphi$, and can
be written as $r(\varphi)=\xi(\varphi)r_0$ where $\xi$ is given as an
analytical function in \citet{Wilkin},
\begin{equation}\label{eq:xi}
\xi(\varphi)=\csc(\varphi)\sqrt{3\left(1-\varphi\cot{\varphi}\right)}
\end{equation}
The velocity of the gas inside of the bow shock is tangential to the
shock surface and equal to
\begin{equation}\label{eq:vl}
v_l=v_w\frac{
  \sqrt{ 
    (\varphi- \sin{\varphi} \cos{\varphi})^2+(3(1-\varphi\cot{\varphi})-\sin^2{\varphi})^2}}
{2(1-\cos{\varphi})+3v_w(1-\varphi\cot{\varphi})/v_*}
\end{equation}
The strength of the outer shock depends on the angle $\varphi$. The
shock will terminate at an angle, $\varphi_{\rm max}$, where the
normal component of the velocity of the ISM with respect to the star,
$v_{*n}$, is equal to the sound speed of the ISM. As in
\citet{Baranov}, it is useful to define the following angles:
$y=\arctan{({\rm d}\ln{\xi}/{\rm d} \varphi)}$, and
$x=y+\pi/2-\varphi$, so that $v_{*n}=v_*\sin{x}$. Then $\varphi_{\rm
  max}$ is found by solving 
\(\mathcal{M}_*\sin{x}=1\). When \(\mathcal{M}_*=2\), $\varphi_{\rm
  max} \approx 106^\circ$ and $\xi_{\rm max}\equiv\xi(\varphi_{\rm
  max})\approx 2.2$. The cylindrical cross-section area of the forward
shock is
\begin{eqnarray}\label{eq:area}
A &=& \pi r_0^2\xi_{\rm max}^2\sin^2{\varphi_{\rm max}}\\
\label{eq:areaEst}
&\sim& 10^{29}\ d_{16}^2
\left(\frac{\dot{M}_*v_w}{2.8\times10^{25}\ {\rm dyn}}\right)
\; {\rm cm^2}.
\end{eqnarray}
$\mathcal{M}_*$ has some degree of uncertainty from 
modeling the Galactic Center, and the fact that the inclination of
G2's trajectory with respect to a possible accretion disk of Sgr A* is
unknown. The change in the area can be estimated using \(\xi_{\rm
  max}^2\sin^2{\varphi_{\rm max}}\approx3.5\mathcal{M}_*-2.4\), when
\(\mathcal{M}_*\geq 1.5\).
The cross-sectional area versus time is plotted in figure
\ref{fig:area}. The area in the wind models decreases by $\sim2$
orders of magnitude as G2 approaches periapsis. Synchrotron flux is
expected to peak at periapsis where the magnetic field is
strongest. The decrease in the area of the forward shock as G2
approaches periapsis is responsible for the large difference in flux
for the pressure-confined cloud and stellar wind models.

\section{Results}\label{sec:RESULTS}

\subsection{Radio Flux of Forward Shock}

To calculate the expected synchrotron emission of the forward shock of
G2 we extend the methodology of \citet{Sadowski} to the stellar wind
model of G2, who used particle-in-cell
  simulations of low Mach number shocks to estimate the electron
  population accelerated by G2. As in \citet{Sadowski}, we assume
that a fraction $\eta \sim 5\%$ of the electrons swept into the
forward shock are accelerated to a high-energy power law spectrum with
a slope of $p=2.4$ and that has a minimum Lorentz factor
$\gamma-1=\zeta kT/m_ec^2$, where $\zeta=7.5$ and $T$ is the preshock
temperature. $\eta$ and $\zeta$ were empirically determined by
\citet{Sadowski} with simulations. They found $\eta$ and $\zeta$ to be
nearly independent of $T$ and $\mathcal{M}_*$. It may be
  counter-intuitive that the injection energy of the power law
  spectrum depends on the preshocked rather than shocked temperature,
  but because of the very narrow range of $\mathcal{M}_*$ applicable
  to G2 and considered by \citet{Sadowski}, {\it i.e.} $1.5\leq
  \mathcal{M}_*\lae 3.5$, it makes little difference which temperature
  one chooses (as long as $\zeta$ is adjusted accordingly).

The energy distribution of the rate at which electrons enter the shock
is given by \citep{Sadowski},
\begin{equation}\label{eq:Cdt}
\frac{{\rm d} N}{{\rm d}\gamma {\rm d}t}=\eta Av_*n
\frac{(p-1)(\zeta k T/m_e c^2)^{p-1}}{(\gamma-1)^{p}}
\end{equation}
for \( \gamma-1> \zeta k T/m_e c^2\), and where $A$, $v_*$, $n$, and
$T$ are all functions of time. To calculate the expected radio flux,
we assume that $P_{\rm mag}=\chi P_{\rm gas}$, where $P_{\rm gas}$ is
the pressure of the unshocked ISM to calculate the unshocked magnetic
field. The shocked magnetic field is calculated assuming only shock
compression with no additional amplification of the magnetic field, as
expected for small Mach number shocks. We use the Rankine-Hugoniot
jump conditions of an adiabatic shock to find the shocked magnetic
field from \citep{Narayan},
\begin{eqnarray}
\label{eq:B}
B &\approx & 0.01\ \chi^{0.5}_{-1}d_{16}^{-1}\ {\rm G}\\ 
\label{eq:Bshocked}
B' & \approx &
\frac{(\hat{\gamma}+1)\mathcal{M}_*^2}{(\hat{\gamma}-1)\mathcal{M}_*^2+2}
B  \sim 0.02\ \chi^{0.5}_{-1}d_{16}^{-1} {\rm G},
\end{eqnarray}
when $\mathcal{M}_*\approx 2$, $\hat{\gamma}=5/3$. 
As an upper limit on $B'$, we make sure that the shocked magnetic
pressure does not exceed the ram pressure of the ISM, this is true for
all Mach numbers if $\chi\leq m_pv_*^2/(16kT)\sim 0.4$.
We take $\chi=0.3$, which corresponds to the trajectory with largest
flux at 1.4 GHz in \citet{Sadowski}. The synchrotron specific
luminosity at the frequency $\nu$ is
\begin{equation}\label{eq:power}
\textstyle
P_\nu(t) = 
\frac{\sqrt{3} q^3 C B}{ m_e c^2 (p+1)}
\Gamma\left(\frac{p}{4}+\frac{19}{12}\right)
\Gamma\left(\frac{p}{4}-\frac{1}{12}\right)
\left(\frac{2\pi m_e c \nu}{3 q B}\right)^{-\frac{p-1}{2}}
\end{equation}
where $C$ is the number of electrons with $\gamma \geq 2$ at time $t$.
The observed flux is calculated using a distance to the Galactic
Center of $d_{A*}=8.33\ {\rm kpc}$:
\begin{equation}
\label{eq:Flux}
F_\nu(t)= \frac{P_\nu(t)}{4\pi d_{\rm A*}^2}
\approx 5 \frac{C_{45}B}{(p+1)}\left(\frac{2\pi m_e c \nu}{3 q B}\right)^{-\frac{p-1}{2}}\ {\rm Jy/G}
\end{equation}


To calculate the flux, we extend the two scenarios defined in
\citet{Sadowski} to a bow shock with changing area which starts
accelerating electrons at sometime $t_i$ and the forward shock reaches
periapsis at $t_0$. The two models are (1) the plow model, where every
electron that enters the shock stays in the shocked area and radiates
in the shocked magnetic field, and (2) the local model, where the
electrons are energized by the shock but then they quickly exit the
shock and radiate \textit{in situ} in the unshocked local magnetic
field.

In the plow model, the flux at 1.4 GHz is calculated from eq
(\ref{eq:Flux}) using the area from eq (\ref{eq:area}) and the shocked
magnetic field, eq (\ref{eq:Bshocked}). 
When $t\leq t_0$, $C_{\rm plow}$ is
\begin{equation}\label{eq:CBefore}
\textstyle
C_{\rm plow}(t)=\eta A (p-1)
\int_{t_i}^{t}
{v_*n\left(\frac{\zeta k T}{m_ec^2}\right)^{p-1} {\ \rm d}t},
\end{equation}
where $A$ is the instantaneous bow shock area at time $t$.
When $t>t_0$, $C_{\rm plow}$ is
\begin{equation}\label{eq:CAfter}
\textstyle
C_{\rm plow}(t)=
C_{\rm plow}(t_0)+
\eta(p-1)
\int_{t_0}^{t}
{Av_*n\left(\frac{\zeta k T}{m_ec^2}\right)^{p-1} {\ \rm d}t}
\end{equation} 

\begin{figure}
\vspace{-.4 cm}
\includegraphics[width=.47\textwidth]{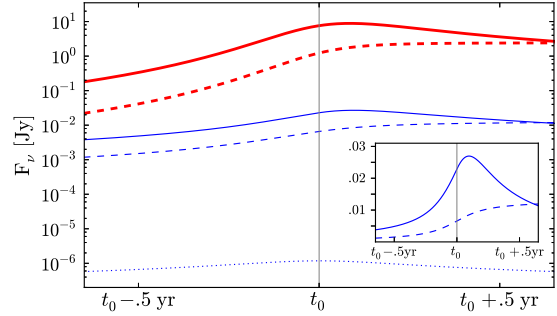}
\caption{The expected synchrotron flux at 1.4 GHz around pericenter
  passage of the forward shock, $t_0$, for different models of
  G2. Solid lines correspond to the plow model, where all of the
  accelerated electrons stay inside of the shock. The dashed lines
  correspond to the local model, where the electrons quickly leave the
  bow shock region after being accelerated. Thick red lines correspond
  to when G2 is a cloud with an area of $\pi \times 10^{30}\ {\rm
    cm^2}$ as in \citet{Narayan,Sadowski}. The lower blue lines
  represent when G2 is a stellar wind with an area calculated using eq
  (\ref{eq:area}). The dotted blue line is the flux from electrons
  residing in the inner shock of the stellar wind assuming a steady
  state solution. {\it Inset}: A plot showing the flux at 1.4
  GHz in Jy for a stellar wind model of G2 in a linear scale.}
\label{fig:lightcurves}
\end{figure}

In the local model, the flux is calculated with the unshocked magnetic
field
\begin{equation}\label{eq:GoodLocal}
F_{{\rm local}}=\frac{5\times10^{-45}}{p+1}\int_{t_i}^t{B
\left(\frac{2\pi m_ec\nu}{3 q B}\right)^{-\frac{p-1}{2}}
\left.\frac{{\rm d} N}{{\rm d}\gamma {\rm d}t}\right|_{\gamma=2}
{\rm d}t} 
\end{equation}
The light curves of the flux at 1.4 GHz in the stellar wind model of
G2 are shown near pericenter passage of the forward shock in the inset
of figure \ref{fig:lightcurves}. The light curves in the stellar wind
model peak shortly after $t_0$ as in the cloud model of
\citet{Sadowski}. To compare the expected fluxes from the stellar wind
and cloud models, we also calculate the expected flux using a constant
area of $A= \pi\times 10^{30}\ {\rm cm^2}$ in equations
(\ref{eq:CBefore}) and (\ref{eq:GoodLocal}). The light curves close to
pericenter of both the cloud and stellar wind models are plotted in
figure \ref{fig:lightcurves}.

To see how our results depend on our parameter choices, we 
estimate $C$ in both the plow and local models by multiplying
$\left.\frac{{\rm d}N}{{\rm d} \gamma {\rm d} t}\right|_{\gamma=2} $
by the dynamical time ($\sim d/v_*$) and using the area from eq
(\ref{eq:areaEst}).
 \begin{eqnarray}
C & \sim & \eta A (p-1)nd(\zeta kT/m_ec^2)^{p-1}\\
\label{eq:C}
& \sim & 2\times 10^{46} d_{16}^{0.6}
\left(\frac{\dot{M}_*v_w}{2.8\times10^{25}\ {\rm dyn}}\right)
\end{eqnarray}
Then the flux at 1.4 GHz when $t\leq t_0$
is estimated using eq (\ref{eq:Flux}) with 
magnetic fields from eq (\ref{eq:B}) and (\ref{eq:Bshocked}) for the
local and plow model respectively
\begin{eqnarray}
\label{eq:LocalEst}
F_{\rm local} & \sim & 0.6\ \left(\frac{\chi}{.3}\right)^{0.85}d_{16}^{-1.1}
\left(\frac{\dot{M}_*v_w}{2.8\times10^{25}\ {\rm dyn}}\right)\; {\rm mJy}\\
 &\sim& 3\ {\rm mJy} \quad {\rm at}\quad t=t_0 \\
\label{eq:PlowEst}
F_{\rm plow} & \sim & 4 \left(\frac{\chi}{.3}\right)^{0.85}d_{16}^{-1.1}
\left(\frac{\dot{M}_*v_w}{2.8\times10^{25}\ {\rm dyn}}\right)\; {\rm mJy}\\
 &\sim& 13\ {\rm mJy} \quad {\rm at}\quad t=t_0
\end{eqnarray}

The analytical results given by equations (\ref{eq:LocalEst}) and
(\ref{eq:PlowEst}) are found to be within a factor of 2 from the
numerical results in figure \ref{fig:lightcurves}.

\subsection{Radio Flux of Inner Shock}
In the stellar wind model of G2, electrons will be accelerated at both
the outer, forward shock traveling into the ISM as well as at the
inner, termination shock of the stellar wind.  
For the inner shock to radiate at 1.4 GHz, it must be able to
accelerate electrons from a upstream temperature of $\sim 1$ eV to
$\sim 3\times10^7$ eV. It is not clear whether a non-relativistic
shock with $\mathcal{M}_w\approx 4$ will be capable of doing this
efficiently. Therefore, our estimation of the radio flux from the
inner shock in this section should be taken as an upper limit. We find
the flux at 1.4 GHz lies four orders of magnitude below the fluxes
calculated in the previous section. Therefore the inner shock will
not contribute to the radio synchrotron flux of G2.

The stellar wind at the bow shock is much more dense than the ISM, its
density equal to
\begin{equation}
n_w(r_0)=\frac{n_{\rm ISM}v_*^2}{v_w^2}
\approx8\times10^{6}\ d_{16}^{-2}\left(\frac{50\ {\rm km/s}}{v_w}\right)^2\ {\rm cm^{-3}}
\end{equation}
The shocked gas will radiate efficiently and cool quickly, and
therefore the shock can be taken to be isothermal. So $n_w$ will be
boosted by a factor $\mathcal{M}_w^2$:
\begin{equation}
n'_w(r_0) = \mathcal{M}_w^2 n_w(r_0) \approx 1.4\times10^{8}\ d_{16}^{-2} T_{w,4}^{-1}\ {\rm cm^{-3}}
\end{equation}
Assuming that in the shocked wind $P'_{\rm mag}= \chi'_w P'_{\rm
  wind}$, $\chi'_w\leq 1$, the magnetic field inside of the inner shock is
\begin{equation}\label{eq:Bboosted}
  B'_{\rm in} = \sqrt{8\pi \chi'_w n'_w(r_0) k T_w}
\approx 0.07\ \chi'^{0.5}_{w} d_{16}^{-1}\ {\rm G} 
\end{equation}

We calculate the volume of the inner shock by assuming that it is in
steady state. In a steady state, the thickness of the shock, $t_l$, is
found by numerically solving the continuity equation for $t_l$ as a
function of $\varphi$,
\begin{equation}\label{eq:ContinuityEq}
2\pi r_0\xi t_l v_l n'_w \sin{\varphi} \approx 2\pi n_w r_0^2\xi^2 v_w
(1-\cos{\varphi}).
\end{equation}
$t_l$ is estimated as being constant in $\varphi$ and equal to
$t_l\sim r_0/\mathcal{M}_w^2$:
\begin{equation}
  t_l\sim 4\times10^{12} d_{16}T_{w,4}
{\textstyle
\left(\frac{\dot{M}/(M_\odot/{\rm yr})}{8.8\times10^{-8}}\right)^{0.5}
\left(\frac{50\ {\rm km/s}}{v_w}\right)^{1.5}}
\ {\rm cm}
\end{equation}
We calculate the volume of the inner shock, $V_{\rm in}$, accounting
for the geometry of the bow shock and solving eq
(\ref{eq:ContinuityEq}) for $t_l$. Approximately, $V_{\rm in}\sim 4\pi
r_0^2 t_l$. Using $V_{\rm in}$ we calculate $C_{\rm in}$, the number of
electrons in the inner shock with $\gamma \geq 2$:
\begin{eqnarray}
\label{eq:C_inner}
C_{\rm in} & =& \eta V_{\rm in}n'_w(p-1)(\zeta k T_w/m_ec^2)^{p-1}\\
&\sim &  5\times10^{41}d_{16}T_{w,4}^{1.4}
{\textstyle
\left(\frac{\dot{M}/(M_\odot/{\rm yr})}{8.8\times10^{-8}}\right)^{1.5}
\left(\frac{50\ {\rm km/s}}{v_w}\right)^{0.5}
}
\end{eqnarray}
The flux is estimated using eq (\ref{eq:Flux}):
\begin{eqnarray}
 F_{\nu} &\sim & 2\times10^{-7} d_{16}^{-0.7}\chi'^{0.85}_wT_{w,4}^{1.4}\\
 & &
\quad\times
\left(\frac{\dot{M}/(M_\odot/{\rm yr})}{8.8\times10^{-8}}\right)^{1.5}
\left(\frac{50\ {\rm km/s}}{v_w}\right)^{0.5}\ {\rm Jy}
\\
&\sim & 0.5\ \mu{\rm Jy} \quad {\rm at} \quad t=t_0
\end{eqnarray}
Using eq (\ref{eq:C_inner}), (\ref{eq:Bboosted}), (\ref{eq:power}) and
(\ref{eq:Flux}), we calculate the expected radio flux for the inner
shock at $1.4\ {\rm GHz}$, and plot it in figure
\ref{fig:lightcurves}. Our estimate underpredicts the more accurate
calculation of the inner shock radio flux at periapsis by a factor $\lae
2$. The radio flux from the inner shock lies well below the forward
shock and can be ignored.

\subsection{Spectra}
\begin{figure}
\vspace{-.4cm}
\includegraphics[width=.47\textwidth]{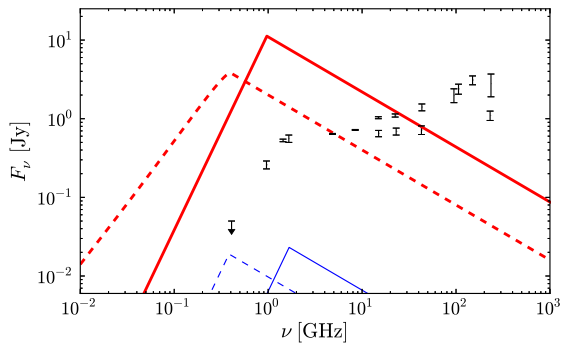}
  \caption{Spectra at a time $t_0+0.05\ {\rm yr}$, where $t_0$ is the
    periapsis crossing time of the forward shock. The color scheme is
    the same as figure \ref{fig:lightcurves}, where solid lines
    correspond to the plow model, and dashed lines to the local
    model. Thick red lines represent the spectra if G2 is a cloud,
    while the lower blue lines are if G2 is a stellar wind. The
    data points are radio fluxes measured during periods of inactivity
    of Sgr A* by \citet{Davies, Falcke, Zhao}.}
\label{fig:spectra}
\end{figure}

In figure \ref{fig:spectra} we show the spectra expected from forward
shock at 0.05 years after periapsis of the forward shock. To calculate
the synchrotron self absorption frequency in the plowing models, we
simply use the surface area of the forward shock at time $t_0+0.05$
years, accounting for the geometry of the bow shock in the stellar
wind model. In the local models, we use the same approach as
\citet{Sadowski}, where synchrotron self absorption frequency and
$F_\nu$ are calculated separately for each time step $\Delta t$ up
until $t_0+0.05$ yr, and then the fluxes are summed. The surface area
of the radiation in the local model is $2\pi r v_*\Delta t$, where
$r=10^{15}$ cm in the cloud model and is $r_0\xi_{\rm
  max}\sin{\varphi_{\rm max}}$ in the stellar wind model. The radio
flux in the stellar wind model lies below the quiescent flux of Sgr A*
at all frequencies and therefore will likely not be observable.

\section{Application to the Star S2}\label{sec:S2}
There is a cluster of young stars at the galactic center, and one of
the brightest of these stars S2, a B type star that has an orbital
period of 15.5 years, reaches nearly twice as close to Sgr A* as G2 (a
distance of $\sim 1.8\times10^{15}\ {\rm cm}$) \citep{S2}. S2 is
expected to have a strong wind which will form a momentum-supported
bow shock similar to G2 in the stellar wind model. The X-ray radiation
from the inner shock was investigated by \citet{GianniosS2}, who
estimated the S2's wind has a mass-loss rate $\dot{M}\sim 10^{-7}
\dot{M}_\odot/{\rm yr}$ and a velocity $\sim1000$ km/s. Using these
parameters, we estimate peak radio flux at periapsis from the
electrons accelerated by the outer shock of S2 using eq
(\ref{eq:PlowEst}). We estimate that the peak flux from the forward
shock of S2 at 1.4 GHz is $\sim.5$ Jy and $\sim .1$ Jy at 14 GHz
(assuming $p=2.4$). The last pericenter crossing of S2 was in 2002.3
and there is good data of Sgr A* around this time. The flux at 15
GHz varies from $\sim$0.8--1.1 Jy and the flux at 23 GHz varies from
$\sim$0.8--1.2 Jy \citep{CmFlares}. While there is variability in the
radio flux before and after the pericenter crossing of S2, there is
not a flare that lasts for a long time or is spectrally consistent
with synchrotron radiation from S2. Throughout the pericenter crossing
time the flux at 23 GHz is equal to or larger than the flux at 15 GHz,
while the opposite would be expected from the synchrotron radiation of
an electron power-law distribution with $p=2.4$. Thus it is likely
this variation is due to intrinsic variation of Sgr A* and not due to
S2. The lack of detection of the forward shock of S2 in 2002.3 gives
us additional confidence in our prediction that the forward-shock of
G2 will not be observable in the radio band.
\section{Conclusions}\label{sec:CONCLUSIONS}
In this paper we calculated the expected radio flux from the forward
shock and reverse shock of G2 assuming that G2 is the inner bow shock
of a stellar wind as suggested by \citet{Scoville, Ballone}. We used
an analytic solution of a momentum-supported bow shock to calculate
the geometry of the forward shock and provide analytical estimates of
the flux at 1.4 GHz at pericenter passage of the forward and reverse
shock to show how the flux depends on the undetermined parameters of
the stellar wind. If G2 is the inner shock of a stellar wind, its
forward shock cylindrical cross section will decrease as G2 approaches
Sgr A* and the ram pressure increases, driving the stagnation point to
a smaller distance from the star. The decrease in area by over two
orders of magnitude of results in a similar decrease in the radio flux
expected at periapsis, falling well below the previous estimates of
$\sim$1-20 Jy predicted by \citet{Narayan, Sadowski}, where they
assumed that G2 was a gas cloud that formed in pressure equilibrium
and that the cross section area stayed constant. If G2 is the inner
bow shock from a stellar wind, the radio flux from the forward shock
lies below the quiescent radio emission of Sgr A* at all frequencies,
so it will be difficult to detect.

If G2 is a pressure-supported gas cloud, it will likely be destroyed
during pericenter passage; on the other hand G2 will survive if it is
the inner bow shock of a stellar wind. Therefore whether or not G2
survives will be important in determining the makeup of G2. However,
the radio emission will peak when the forward shock crosses periapsis,
which in the cloud model is predicted to happen 7 to 9 months before
G2 passes through periapsis. We have shown that the stellar wind model
makes a very different prediction of the radio synchrotron flux of G2,
so the radio flux will be an important early clue about the nature of
G2.

The authors would like to thank John Lacy for useful discussions, and
Stefan Gillessen, Ramesh Narayan and Roberto Hern{\'a}ndez for their
helpful comments.

\footnotesize{
\bibliographystyle{mn2e}
\bibliography{g2.bib}
}
\end{document}